\documentclass[pra,aps,twocolumn]{revtex4}
\usepackage{amsfonts}
\usepackage{amsmath}
\usepackage{amssymb}
\usepackage{charter}
\usepackage{graphicx}
\usepackage{float}
\usepackage{amsmath}
\usepackage{amssymb}
\usepackage{charter}
\usepackage{graphicx}
\usepackage{subfigure}
\usepackage{graphicx}
\usepackage{epstopdf}
\usepackage{color}
\usepackage{tocvsec2}
\usepackage{enumerate}
\usepackage{graphicx}
\usepackage{dcolumn}
\usepackage{bm}
\usepackage[colorlinks, linkcolor=blue, citecolor=blue, urlcolor=blue, breaklinks=true]{hyperref}

\begin{document}

\title{ Phase estimation via delocalized photon subtraction operation inside
the SU(1,1) interferometer }
\author{Zhihao Li$^{1}$}
\author{Qingqian Kang$^{1,2}$}
\author{Teng Zhao$^{1}$}
\author{Cunjin Liu$^{1}$}
\author{Liyun Hu$^{1,*}$}
\author{Chengzhi Deng$^{3,}$}
\thanks{Corresponding authors: hlyun@jxnu.edu.cn, dengcz@nit.edu.cn}
\affiliation{$^{{\small 1}}$\textit{Center for Quantum Science and Technology, Jiangxi
Normal University, Nanchang 330022, China}\\
$^{{\small 2}}$\textit{Department of Physics, Jiangxi Normal University
Science and Technology College, Nanchang 330022, China}\\
$^{{\small 3}}$\textit{School of information engineering, Nanchang institute of technology, Nanchang 330099, China}}
\begin{abstract}
We propose a theoretical scheme to improve the precision of phase
measurement using intensity detection by implementing delocalized photon
subtraction operation (D-PSO) inside the SU(1,1) interferometer, with the
coherent state and the vacuum state as the input states. We compare the
phase sensitivity and the quantum Fisher information between D-PSO and
localized photon subtraction operation (L-PSO) under both ideal and
photon-loss cases. It has been found that the D-PSO can improve the
measurement accuracy of the SU(1,1) interferometer and enhance its
robustness against internal photon loss. And it can cover and even exceed
the advantages of the L-PSO on two modes, respectively. In addition, by
comparing the standard quantum limit, the Heisenberg limit and quantum Cram%
\'{e}r-Rao bound, we find that the phase sensitivity of the D-PSO can get
closer to the quantum Cram\'{e}r-Rao bound and has the ability to resist
internal loss.

\end{abstract}

\maketitle

\section{Introduction}

With the rapid development of quantum information field, quantum metrology
has been widely studied in recent years \textbf{\cite{c1,c2,c3,c4,c5}}.
Quantum metrology has important applications in many fields, such as
gravitational detection \textbf{\cite{c6,c7,c8,c9}}, high-precision time
measurement \textbf{\cite{c10,c11,c12,c13}}, and quantum imaging \textbf{%
\cite{c14,c15,c16,c17,c18}}. Quantum precision measurement is a branch of
quantum metrology, which leverages the principles of quantum mechanics, such
as quantum superposition and quantum entanglement, to enhance the
measurement accuracy. It's purpose is to break through the precision limits
of classical measurement and provide technical support for quantum computing
and quantum communication. Since quantum entanglement can defeat the noise
in the process of measurement \textbf{\cite{a1}, }the design of sensors
based on the principles of quantum mechanics can make the measurements more
precise compared to that based on the principles of classical physics.

Optical interferometers used for measuring small phase shifts are important
tools for achieving high precision in quantum metrology, among which
Mach-Zehnder interferometer (MZI) has been used as a general model for
achieving precise phase measurement. It uses classical coherent states as
the inputs and can achieve an accuracy of $\Delta \phi =\frac{1}{\sqrt{N}}$,
where $N$ is the average number of photons within the interferometer \textbf{%
\cite{a2}}. To satisfy the need for high precision, people continue to make
improvements to the interferometer: by using non-classical states such as
NOON state \textbf{\cite{a3}}, Fock states \textbf{\cite{a4}}, squeezed
states \textbf{\cite{a5,b1}}, Gaussian states \textbf{\cite{a6} }as inputs
of the MZI and using unbalanced MZI \textbf{\cite{a7}, }its phase
sensitivity can better be than the standard quantum limit (SQL) $\Delta \phi
_{SQL}=\frac{1}{\sqrt{N}}$, and even surpass the Heisenberg limit (HL) $%
\Delta \phi _{HL}=\frac{1}{N}$ \textbf{\cite{b2}}. In addition, nonlinear
active elements have also been introduced, replacing traditional 50:50
linear beam splitters (BSs) with active nonliner optical devices such as
optical parametric amplifiers (OPAs) and four-wave mixers (FWMs), which use
fewer optical elements than the SU(2) interferometers and hence present a
more practical way of making sensitive interferometry. This kind of
interferometer was proposed by Yurke \textit{et al.}\textbf{\cite{a8}}.
Owing to that the second nonlinear element annihilates some of the photons
inside, Zhang \textit{et al.}\textbf{\cite{a9}} proposed a different
protocol based on a modified SU(1,1) interferometer, where the second
nonlinear element is replaced by a beam splitter, which can achieve the
sub-shot-noise-limited phase sensitivity and is robust against photon loss
as well as background noise.

In addition to improving the interferometer itself, people have also made
improvements to the input photon states. For example, Zhang \textit{et al.}%
\textbf{\cite{a10}} introduced the multiphoton catalysis two-mode squeezed
vacuum state as an input of the MZI and studied its phase sensitivity with
photon-number parity measurement and the influence of photon loss. Kang 
\textit{et al.}\textbf{\cite{a11}} proposed a theoretical scheme to improve
the precision of phase measurement using homodyne detection by implementing
a number-conserving operation, inside the SU(1,1) interferometer, with the
coherent state and the vacuum state as the input states. Besides, a scheme
of subtracting one photon from the output state of an SU(1,1) interferometer
had been put forward, which can not only improve the output state but also
be used to optimize the encoding procedure \textbf{\cite{a12}}. These
non-Gaussian operations can generate non-Gaussian states \textbf{\cite{b3}}
and therefore improve the phase sensitivity or eliminate the influence of
noise. The above are all schemes for improving the measurement accuracy
based on localized operations acting on the states at the input ports,
output ports and intermediate ports of the interferometers \textbf{\cite%
{b4,b5,b6,b7,b8}}.

Except for localized non-Gaussian operations, in recent years, delocalized
non-Gaussian operations have also drawn extensive attention in the academic
community, which have been applied to improve non-classicality, quantum
entanglement and remote phase sensing. For example, Sun \textit{et al.}%
\textbf{\cite{b9}} use nonlocality to enhance the detection of quantum
entanglement, demonstrating how to efficiently derive a broad class of
inequalities for entanglement detection in multi-mode continuous variable
systems. Biagi \textit{et al.}\textbf{\cite{a13} }developed a method based
on the delocalized addition of a single photon, besides allowing one to
entangle arbitrary states of arbitrarily large size, can generate
discorrelated states. And by exploring the case of delocalized photon
addition over two modes containing identical coherent states, they also
derive the optimal observable to perform remote phase estimation from
heralded quadrature measurements \textbf{\cite{a15}}. In addition, they
realized the delocalized non-Gaussian operation exprimentally, imparting
practical significance to our research. This research experimentally
establishes and measures significant entanglement based on the delocalized
heralded addition of a single photon, between two identical weak laser
pulses containing up to 60 photons each \textbf{\cite{f1}}.

So, will delocalized non-Gaussian operations also improve the measurement
accuracy? And what are the differences in the improvement effects on quantum
precision measurement between delocalized non-Gaussian operations and
localized ones? As far as we know, there has been no related work proposing
that delocalized photon subtraction operations (D-PSO) can improve phase
sensitivity, and there have been relatively few comparisons with localized
non-Gaussian operations. Therefore, we propose to use D-PSO inside the
SU(1,1) interferometer to improve its measurement accuracy, and compare it
with the localized photon subtraction operation (L-PSO) scheme. At the same
time, we analyze the influence of D-PSO on the quantum Fisher information
(QFI) in the presence of photon loss and expect that delocalized
non-Gaussian operations will be able to demonstrate their advantages.

In our research, we find that D-PSO can indeed improve the measurement
accuracy more effectively. The organization of the remaining part of this
paper is as follows. Sec. II outlines the theoretical model of D-PSO. Sec.
III studies the phase sensitivity in both ideal and lossy cases. Sec. IV
investigates the QFI in the ideal and lossy cases and compare the phase
sensitivities with theoretical limits. Finally, a comprehensive summary is
provided.

\section{\protect\bigskip model}

In this section, we first introduce the model of D-PSO inside the SU(1,1)
interferometer. The SU(1,1) interferometer is typically composed of two OPAs
and a linear phase shifter. The first OPA is characterized by a two-mode
squeezing operator $U_{S_{1}}(\xi _{1})=\exp (\xi _{1}^{\ast }ab-\xi
_{1}a^{\dagger }b^{\dagger })$, where $a$ and $b$ are the annihilation
operators for the two modes, respectively. The squeezing parameter $\xi _{1}$
can be expressed as $\xi _{1}=g_{1}e^{i\theta _{1}}$, where $g_{1}$
represents the gain factor and $\theta _{1}$ represents the phase shift. The
input ports are $\left\vert \alpha \right\rangle _{a}\otimes \left\vert
0\right\rangle _{b}$, where $\left\vert \alpha \right\rangle _{a}\ $is
coherent state and $\left\vert 0\right\rangle _{b}$ is vacuum state. We
perform a D-PSO behind the first OPA, which can be expressed as $\left(
sa+tb\right) ^{m}$, where $s$ and $t$ are the proportionality coefficient of
mode $a$ and mode $b$ in the operation, respectively. They are real numbers. 
$m$ represents the order. Since the SU(1,1) interferometer is prone to
internal photon loss in real situations, we use two fictitious BSs behind
the D-PSO to simulate photon loss. The operators of these fictitious BSs can
be represented as $U_{B_{a}}=\exp \left[ \left( a^{\dagger
}a_{v}-aa_{v}^{\dagger }\right) \arccos \sqrt{T_{a}}\right] $ and $%
U_{B_{b}}=\exp \left[ \left( b^{\dagger }b_{v}-bb_{v}^{\dagger }\right)
\arccos \sqrt{T_{b}}\right] $, $T_{a}$ and $T_{b}$ are the transmissivity of
the fictitious BSs. For simplicity, we take $T_{a}=T_{b}=T$. $T=1$
corresponds to the ideal situation, that is, there is no photon loss.
Following the first OPA, mode $a$ undergoes a phase shift process $U_{\phi
}=\exp [i\phi (a^{\dagger }a)]$ afterwards, but mode $b$ remains unchanged.
They ultimately passes through the second OPA $U_{S_{2}}(\xi _{2})=\exp (\xi
_{2}^{\ast }ab-\xi _{2}a^{\dagger }b^{\dagger })$, where $\xi
_{2}=g_{2}e^{i\theta _{2}}$. At the same time, performing intensity
detection on mode $a$ and mode $b$. The OPAs process can be equivalent to
the process of two-mode squeezing, and a balanced situation of the two OPAs
is $\theta _{2}-\theta _{1}=\pi $ and $g_{1}=g_{2}$. In this paper, we set $%
g_{1}=g_{2}=g$, $\theta _{1}=0$, $\theta _{2}=\pi $. After the influence of
noise, the quantum state will become a mixed state. However, if it is
described in the extended space, the expression for the output state of the
standard SU(1,1) interferometer can be represented as the following pure
state: 
\begin{equation}
\left\vert \Psi _{out}\right\rangle=AU_{S_{2}}U_{\phi }U_{B}\left(
sa+tb\right) ^{m}U_{S_{1}}\left\vert \Psi _{in}\right\rangle,  \label{1}
\end{equation}%
where $\left\vert \Psi _{in}\right\rangle =\left\vert \alpha \right\rangle
_{a}\left\vert 0\right\rangle _{b}\left\vert 0\right\rangle _{av}\left\vert
0\right\rangle _{bv}$, $s+t=1(0\leq s\leq 1)$. $s=1$, $t=0$ equivals to
performing a L-PSO only on mode $a$, and $s=0$, $t=1$ equivals to performing
a L-PSO only on mode $b$. Others are all D-PSO.

\begin{figure}[tph]
\label{Figure1} \centering\includegraphics[width=0.95%
\columnwidth]{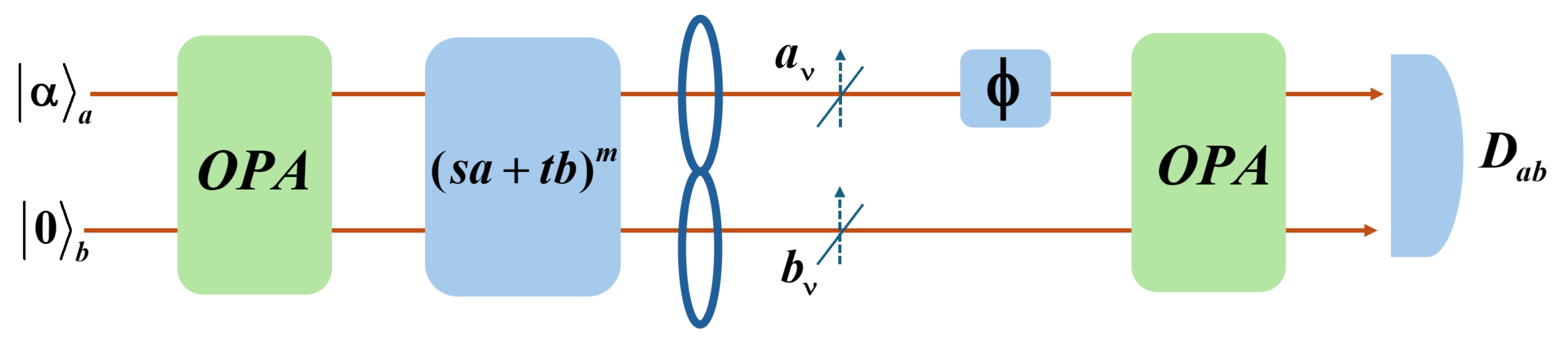} \ 
\caption{Schematic diagram of D-PSO within the SU(1,1). OPA is the optical
parametric amplifier, D-PSO act behind the first OPA. The symbol of $\protect%
\phi $ is corresponding to the phase shifter, and $D_{ab}$ is the intensity
detector between mode $a$ and mode $b$. The two fictitious BSs inside the SU
(1,1) interferometer are used to describe the photon losses on the mode $a$
and mode $b$ , respectively.}
\end{figure}
The normalization constants can be calculated as: 
\begin{equation}
A=\left( Q_{m,0,0,0,0}\right) ^{-\frac{1}{2}}.  \label{2}
\end{equation}
The specific derivation of $Q_{m,x_{1},y_{1},x_{2},y_{2}}$ is shown in
Appendix A.

\section{Phase sensitivity}

In phase estimation, different detection methods can lead to different
effects on phase sensitivity. Selecting an appropriate detection method can
measure phase changes more accurately and minimize errors. Commonly used
detection methods are: intensity detection \textbf{\cite{a16,a17}}, homodyne
detection \textbf{\cite{a18,a19}}, parity detection \textbf{\cite{a20,a21}}.
Considering the accuracy and simplicity of intensity detection, in our
scheme, we perform intensity detection on mode $a$ and $b$.

The photon-number sum operator of mode $a$ and mode $b$ is $X=a^{\dagger
}a+b^{\dag }b$, based on the error-propagation equation \textbf{\cite{a8}},
the phase sensitivity of the SU(1,1) interferometer can be given by the
following formula: 
\begin{equation}
\Delta \phi =\frac{\sqrt{\left\langle \Delta ^{2}X\right\rangle }}{|\partial
\left\langle X\right\rangle /\partial \phi |}=\frac{\sqrt{\left\langle
X^{2}\right\rangle -\left\langle X\right\rangle ^{2}}}{|\partial
\left\langle X\right\rangle /\partial \phi |},  \label{7}
\end{equation}%
where $\langle X\rangle =\left\langle \Psi _{out}\right\vert (a^{\dagger
}a+b^{\dag }b)\left\vert \Psi _{out}\right\rangle$. Detailed calculation
steps for the phase sensitivity $\Delta \phi $ of the D-PSO are provided in
Appendix A.

First, we consider the effects of different parameters on the phase
sensitivity without photon loss. We take the influence of the order $m$ on
the phase sensitivity into account, where $T=1$ representing the scenario
without photon losses. We fix $\alpha =1$, $g=1$ and the optimal phase
sensitivity is obtained by optimizing the coefficient $t$ of the D-PSO. We
compare the influences of without non-Gaussian operation, L-PSO\ and D-PSO
on the phase sensitivity respectively in Fig. 2. It is shown that, (i) when
there is no non-Gaussian operation, the phase sensitivity $\Delta \phi $
first improves and then weakens with the increase of phase shift $\varphi \ $%
and the phase sensitivity reaches its optimal value $\Delta \phi \approx
0.25 $ when $\left\vert \varphi \right\vert \approx 0.7$.$\ $(The smaller
the phase sensitivity is, the higher the precision is.) (ii) In the presence
of L-PSO, the variation trend of phase sensitivity $\Delta \phi $ with phase
shift $\varphi $ is the same as described above. Notably, the value of the
optimal phase sensitivity $(\Delta \phi \approx 0.18)$ outperforms that in
the previous case, while the absolute value of the corresponding phase shift
is larger. That is, the value of $\varphi $ corresponding to the optimal
point shifts toward larger values of $\varphi $. Additionally, the phase
sensitivity of mode $b$ is superior to that of mode $a$. (iii) It is
interesting that, the optimal range of phase shift $\varphi $\ in D-PSO can
encompass the optimal ranges of both modes $a$ and $b$, which means that the
D-PSO is able to combine advantages of the L-PSO. And in almost the entire
space of $\varphi $, D-PSO demonstrates the best performance compared with
the original operation and the L-PSO scheme. (iv) As the order $m$
increases, the optimal phase sensitivities $\Delta \phi $\ of the two L-PSOs
gradually improve, and the difference in their corresponding phase shifts $%
\varphi $ gradually increases. (Combined with the Fig. 2, the phase
sensitivity $\Delta \phi $\ reaches its optimum when $\varphi $\ is
approximately equal to 1. So in order to facilitate comparison, we take $%
\varphi =1$\ in the subsequent images.) (v) Compared with the standard case,
both L-PSO and D-PSO can improve the phase sensitivity and D-PSO has a
better improvement effect than L-PSO. Therefore, in the following sections,
we will only focus on the comparison between L-PSO and D-PSO. 
\begin{figure}[tbp]
\label{Figure2} \centering\includegraphics[width=0.95%
\columnwidth]{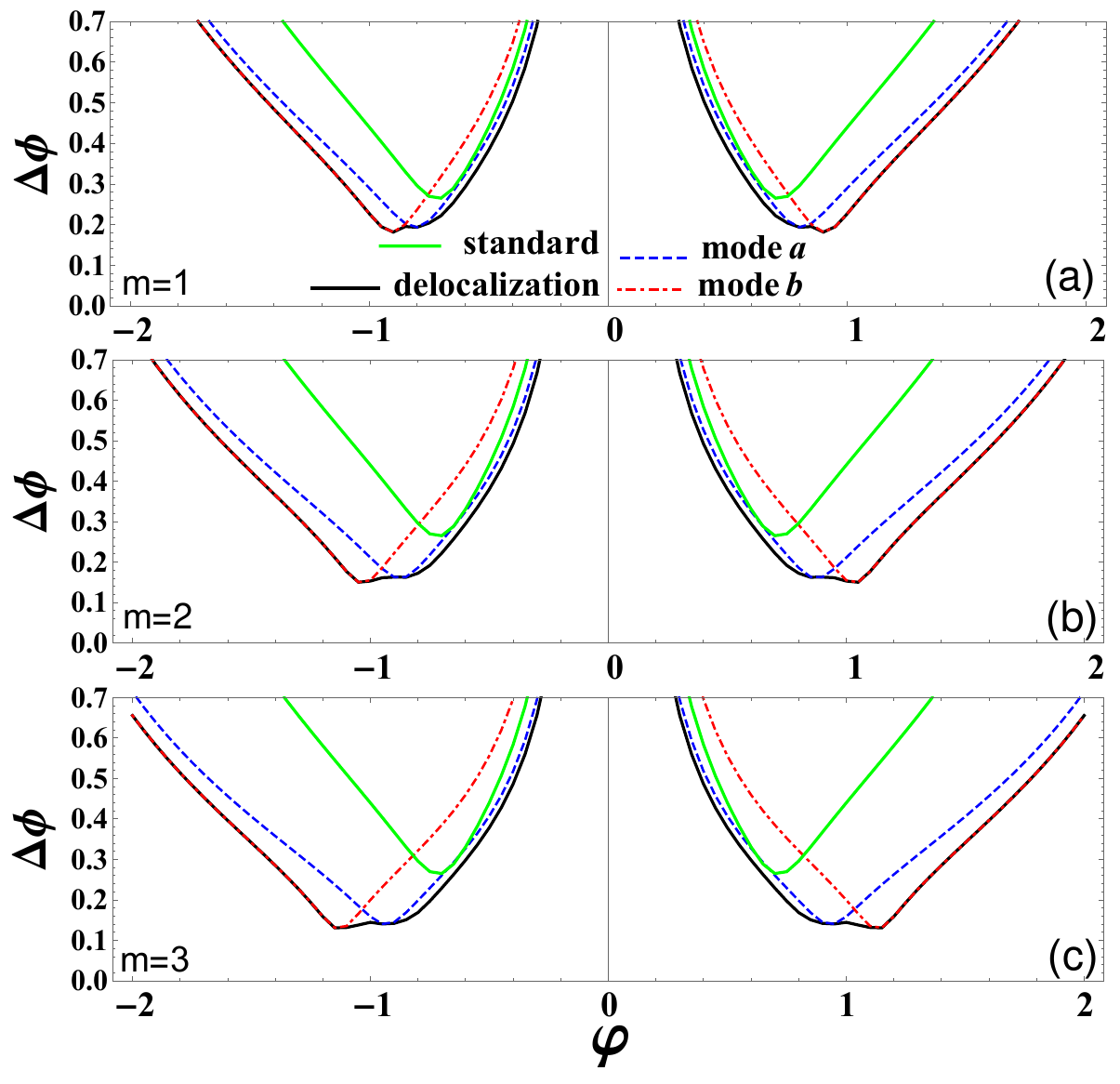} \ 
\caption{The comparison between D-PSO and L-PSO, about phase sensitivity $%
\Delta \protect\phi $ as a function of $\protect\varphi $, with $\protect%
\alpha =1$, $g=1\ $and$\ T=1$. Standard represents the standard SU(1,1)
interferometer (without non-Gaussian operations). Mode $a$ represents the
photon subtraction operation only performed on mode $a$ $(s=1,t=0)$, mode $b$
represents the photon subtraction operation only performed on mode $b$ $%
(s=0,t=1)$ and delocalization means D-PSO.}
\end{figure}
\begin{figure}[tph]
\label{Figure3} \centering\includegraphics[width=0.95%
\columnwidth]{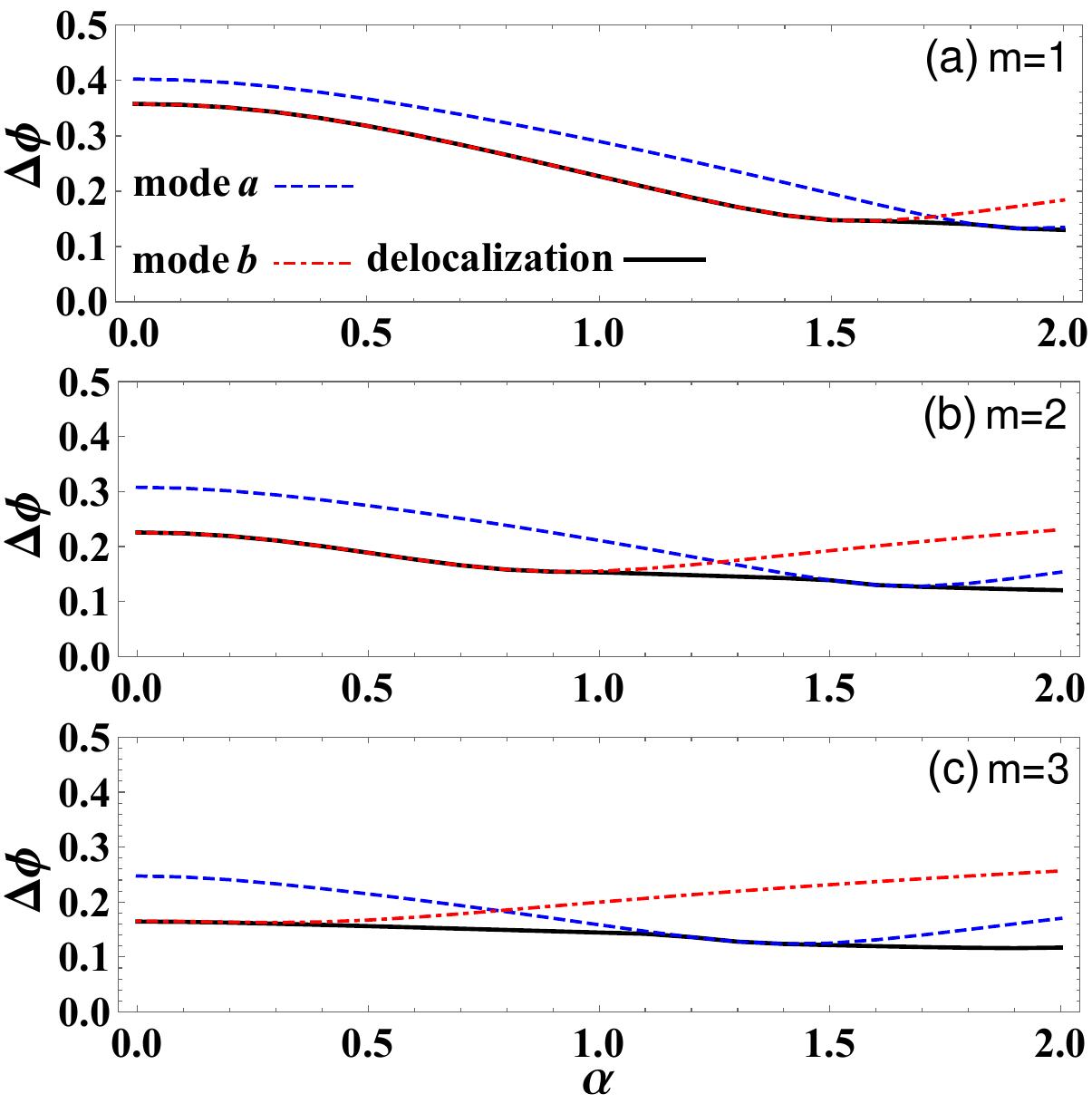} \ 
\caption{The comparison between D-PSO and L-PSO, about phase sensitivity $%
\Delta \protect\phi $ as a function of coherent amplitude $\protect\alpha $,
with $\protect\varphi =1$, $g=1\ $and$\ T=1$.}
\end{figure}
\begin{figure}[tbp]
\label{Figure4} \centering\includegraphics[width=0.95%
\columnwidth]{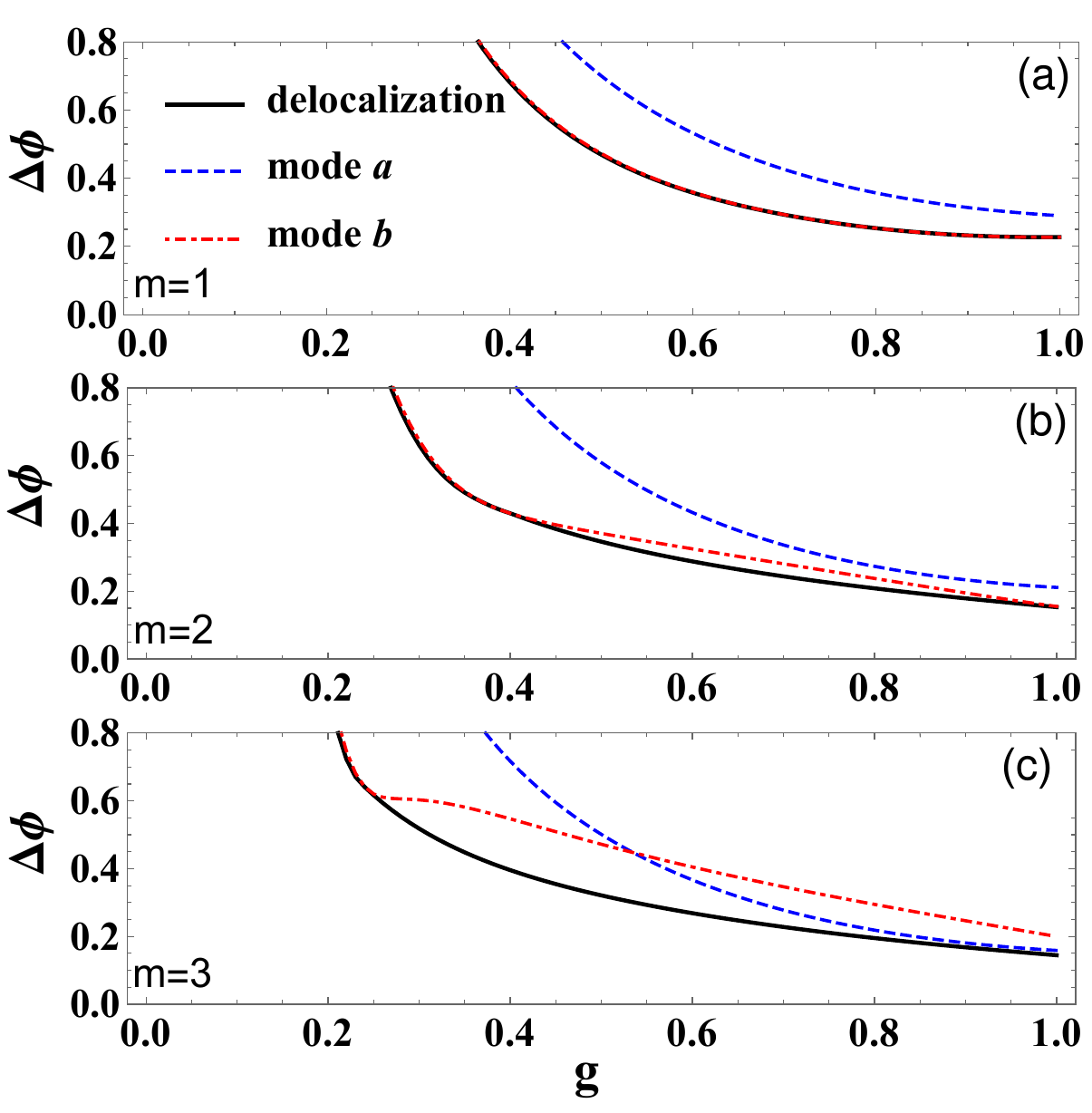} \ 
\caption{The comparison between D-PSO and L-PSO, about phase sensitivity $%
\Delta \protect\phi $ as a function of gain factor $g$, with $\protect%
\varphi =1$, $\protect\alpha =1\ $and$\ T=1$.}
\end{figure}

Next, we further investigate the influence of other parameters on the phase
sensitivity in ideal case. The phase sensitivity $\Delta \phi $\ is plotted
in Fig. 3 as a function of coherent amplitude $\alpha $. By comparing the
effects of L-PSO\ and D-PSO, it is interesting to notice that, (i) with the
increasing of coherent amplitude $\alpha $, phase sensitivity $\Delta \phi $%
\ gardually improves and then decreases when L-PSO are applied. And when $%
\alpha <1.75$ and $m=1$, the phase sensitivity $\Delta \phi $ of mode $b$ is
greater than that of mode $a$. But as $m$ increases, the range of $\alpha $
for which mode $b$ outperforms mode $a$ will gradually decrease. Besides,
the coherent amplitude $\alpha $\ value corresponding to the optimal phase
sensitivity $\Delta \phi $ of mode $a$ $(\alpha \approx 1.5)\ $is bigger
than that of mode $b$ $(\alpha \approx 1.9)$ and the difference in coherent
amplitudes between the two L-PSOs increases with the order $m$ grows. (ii)
The phase sensitivity $\Delta \phi $\ of the D-PSO operation can encompass
the optimal phase sensitivity $\Delta \phi $ of the L-PSO operation and it
always maintains a low trend almost without rising. So when the value of the
order $m$ increases, the range of coherent amplitude $\alpha $ within which
the phase sensitivity of D-PSO outperforms that of L-PSO gradually expands.

Fig. 4 shows the curve of how the phase sensitivity $\Delta \phi $\ varies
with gain factor $g$. We find that, (i) the phase sensitivity $\Delta \phi $
improves as gain factor$\ g$ increases under the condition of L-PSO. (ii)
The phase sensitivity $\Delta \phi $\ of mode $b$ is generally superior to
that of mode $a$, but this trend undergoes a change when $m=3$ and $g>0.55 $%
. (iii) D-PSO can encompass the optimal phase sensitivities in the two
L-PSOs, and when there is an increase in the order $m$, the advantages of
D-PSO over L-PSO in terms of phase sensitivity gradually emerge. 
\begin{figure}[th]
\label{Figure5} \centering\includegraphics[width=0.95%
\columnwidth]{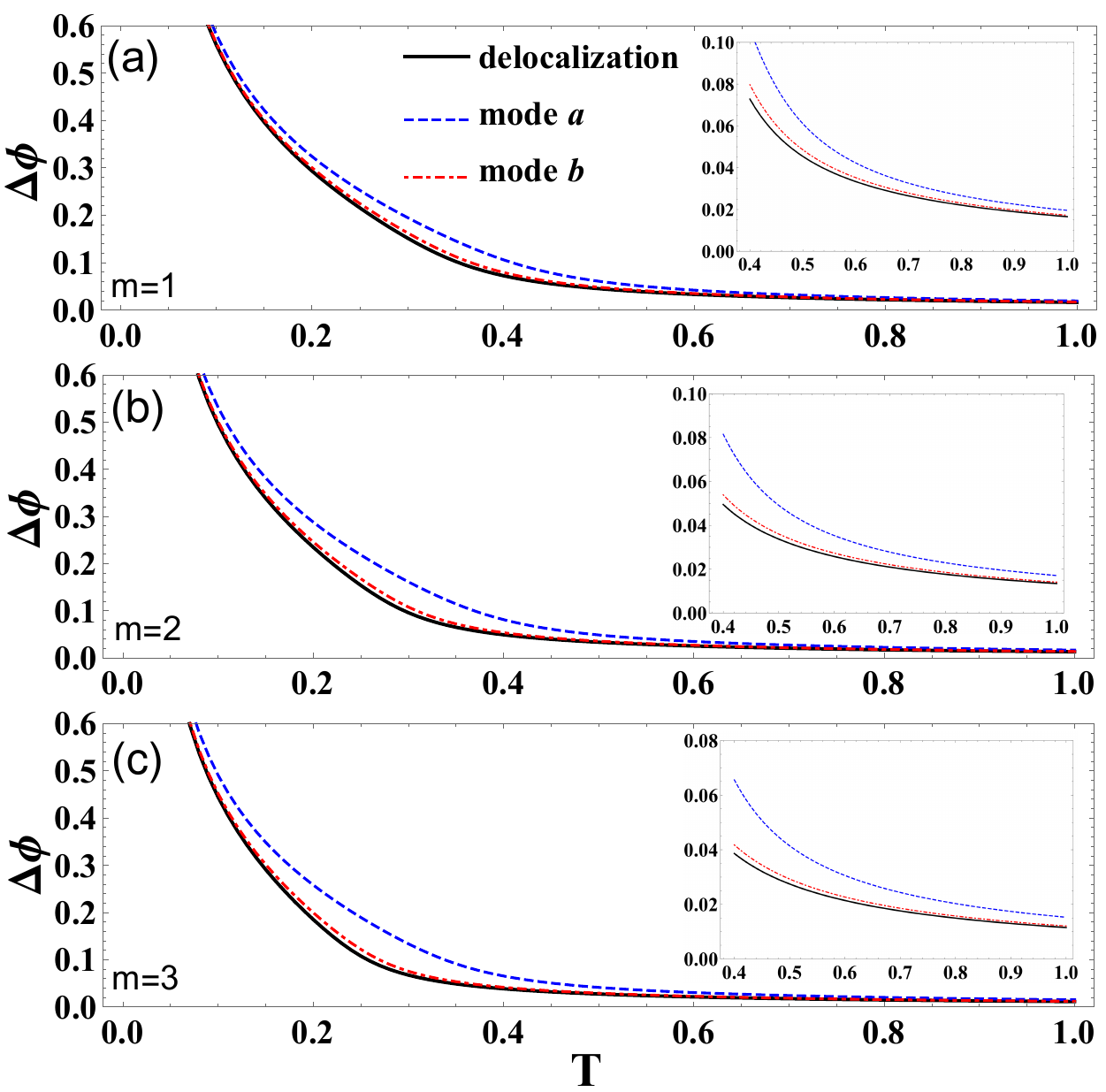} \ \vspace{0.2cm}
\caption{The comparison between D-PSO and L-PSO, about phase sensitivity $%
\Delta \protect\phi $ as a function of $T$, with $\protect\varphi =1$, $g=1\ 
$and$\ \protect\alpha =3$.}
\end{figure}

Since there exists loss in practical situations, now we consider the phase
sensitivity $\Delta \phi $ in the case of loss. Here, $T$ is the
transmissivity of the two fictitious BSs, and $T=1$ and $T=0$ represent the
ideal and completely lossy cases. It is found from Fig. 4 as follows, (i) in
most of the $T$ range ( $T$ is in the range of 0.4 to 1), both L-PSO and
D-PSO show little variation with $T$, indicating that the two schemes are
less affected by loss. Additionally, the sensitivity of mode $b$ in L-PSO outperforms that of mode $a$, and an increase in the order $m$ results in
improved sensitivity. (ii) Throughout the entire $T$ range, the sensitivity
of D-PSO is better than that of L-PSO, indicating that D-PSO exhibits
stronger robustness against photon loss over a wide range.

\section{\protect\bigskip QFI and some theoretical limits}

\subsection{QFI in ideal case}

One of the popular ways to obtain useful bounds in quantum metrology,
without the need for cumbersome optimization, is to use the concept of the
QFI, which represents the maximum information extracted from the
interferometer system \textbf{\cite{a22}.}

For a pure-state system, the QFI can be derived by \textbf{\cite{a23}} 
\begin{equation}
F=4\left[ \left\langle \Psi _{\phi }^{\prime }|\Psi _{\phi }^{\prime
}\right\rangle -\left\vert \left\langle \Psi _{\phi }^{\prime }|\Psi _{\phi
}\right\rangle \right\vert ^{2}\right] ,  \label{9}
\end{equation}%
where\ $\left\vert \Psi _{\phi }\right\rangle $ is the quantum state after
the phase shift and before the second OPA and $\left\vert \Psi _{\phi
}^{\prime }\right\rangle =\partial \left\vert \Psi _{\phi }\right\rangle
/\partial \phi $. Then the QFI can be rewritten as \textbf{\cite{a23}}: 
\begin{equation}
F=4\left\langle \Delta ^{2}n_{a}\right\rangle ,  \label{10}
\end{equation}%
where $\left\langle \Delta ^{2}n_{a}\right\rangle =\left\langle \Psi _{\phi
}\right\vert (a^{\dagger }a)^{2}|\Psi _{\phi }\rangle -(\left\langle \Psi
_{\phi }\right\vert a^{\dagger }a|\Psi _{\phi }\rangle )^{2}$. \bigskip In
our scheme, the quantum state is given by: 
\begin{equation}
\left\vert \Psi _{\phi }\right\rangle =AU_{\phi }\left( sa+tb\right)
^{m}U_{S_{1}}|\psi _{in}\rangle ,\mathit{\ }
\end{equation}
where\textit{\ }$|\psi _{in}\rangle =\left\vert \alpha \right\rangle
_{a}\otimes \left\vert 0\right\rangle _{b}$. Thus, the QFI is derived as: 
\begin{equation}
F=4\left[ A^{2}\left( Q_{2,2,0,0}+Q_{1,1,0,0}\right) -(A^{2}Q_{1,1,0,0})^{2}%
\right] .  \label{11}
\end{equation}

\begin{figure}[tbp]
\label{Figure6} \centering\includegraphics[width=0.95%
\columnwidth]{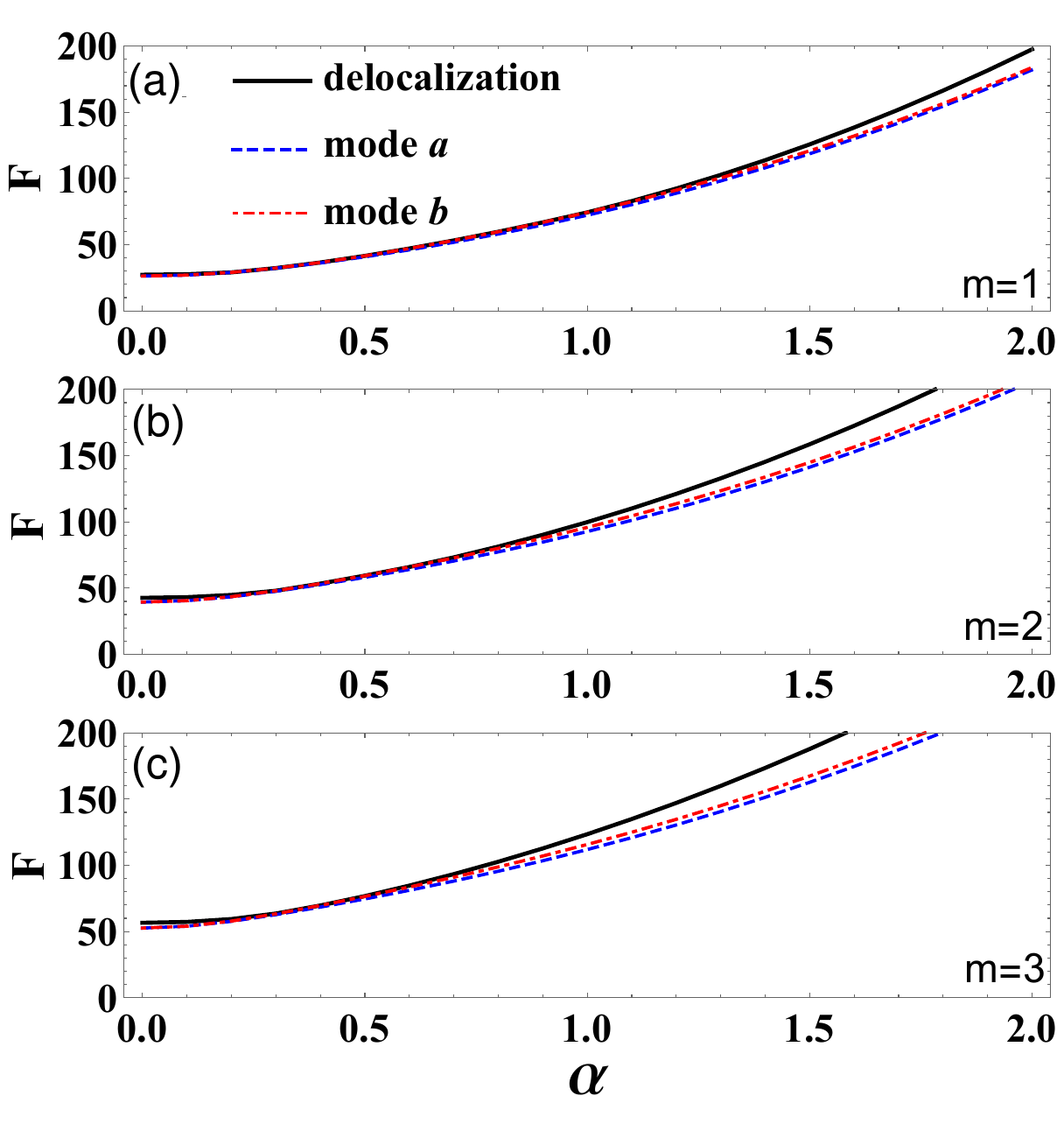}\ 
\caption{The comparison between D-PSO and L-PSO, on QFI as a function of $%
\protect\alpha $, with $\protect\varphi =1$, $g=1\ $and$\ T=1$.}
\end{figure}

In order to investigate how the QFI varies with the parameters, we plot the
curves of the QFI as a function of coherent amplitude $\alpha $ and gain
factor$\ g$. In Fig. 6, it is clear to see that, (i) under a given $m$, with
the increasing of coherent amplitude $\alpha $, the QFI of both D-PSO and
L-PSO will enlarge. (ii) Compared to the L-PSO, the QFI of D-PSO will be
larger when the value of $\alpha $ is greater than 1 ($m=1$), which will be
more obvious as the order $m$ grows.

\begin{figure}[tph]
\label{Figure7} \centering\includegraphics[width=0.95%
\columnwidth]{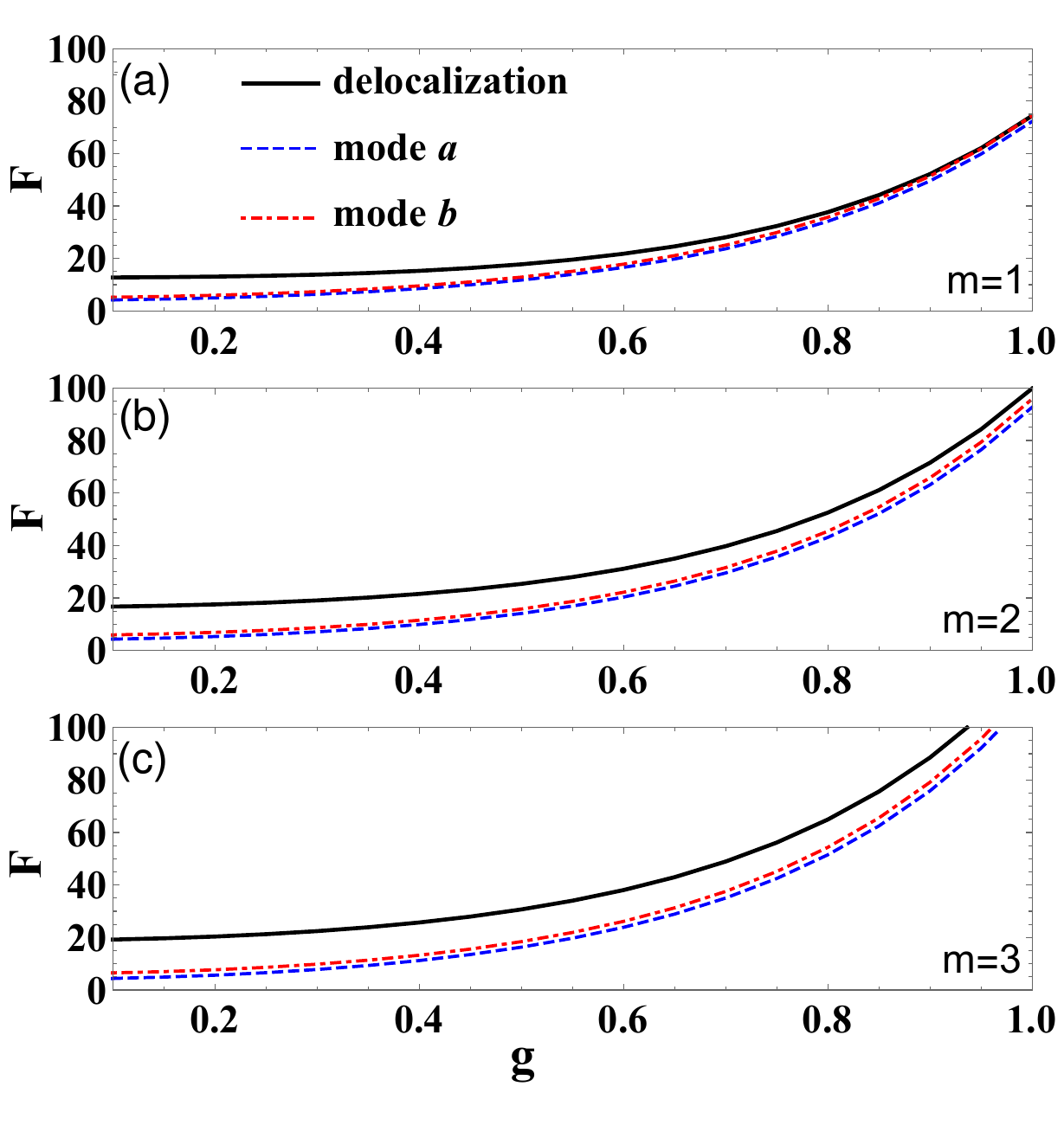} \ 
\caption{The comparison between D-PSO and L-PSO, on QFI as a function of $g$%
, with $\protect\varphi =1$, $\protect\alpha =1\ $and$\ T=1$.}
\end{figure}

Fig. 7 shows that the QFI as a function of $g$, with $\alpha =1$, $T=1$. We
notice that, (i) the QFI goes up as gain factor\ $g$ gets larger. (ii) When
the order $m$ is the same, it is clearly found that the QFI of D-PSO is
greater than that of L-PSO. (iii) As the order $m$ increases, the difference
of QFI between the two L-PSOs becomes more pronounced, and the discrepancy
in QFI between the D-PSO and L-PSO also becomes more evident.

\begin{figure}[tph]
\label{Figure8} \centering\includegraphics[width=0.95%
\columnwidth]{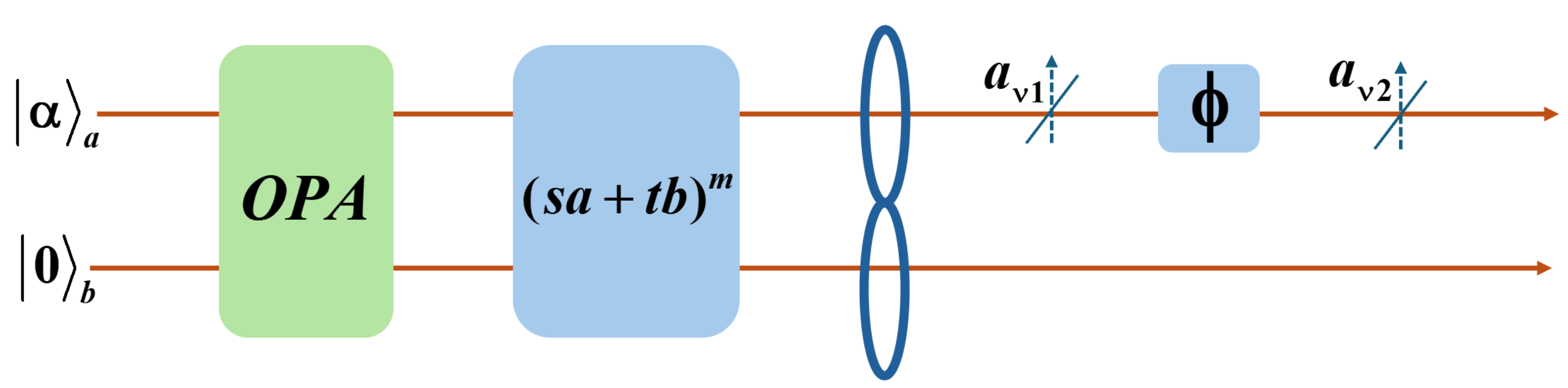} \ 
\caption{The calculation model of the Fisher information under losses. The
loss occurs on the mode $a$.}
\end{figure}

\subsection{\protect\bigskip QFI in lossy case}

In practical situations, the influence of photon loss on the phase\
estimation cannot be neglected. The QFI calculation model under loss
condition is shown in Fig. 8. The small phase shift occurs in mode $a$
inside the SU(1,1) interferometer. For simplicity, we only consider the
photon loss in mode $a$. The method of calculating the QFI in the presence
of photon loss is proposed by Escher \textit{et al. }\textbf{\cite{a23}},
which can be briefly summarized in the following: for an arbitrary initial
pure state $\left\vert \Psi _{s}\right\rangle $, in the probe system S, the
QFI with photon loss can be calculated as: 
\begin{equation}
F_{L}=\min C_{Q}\left[ \left\vert \psi _{S}\right\rangle ,\Pi _{l}\left(
\phi \right) \right] ,  \label{12}
\end{equation}%
where $\Pi _{l}\left( \phi \right) $ is the Kraus operator which acts on the
system S and describes the photon loss, and $C_{Q}\left[ \left\vert \psi
_{S}\right\rangle ,\Pi _{l}\left( \phi \right) \right] $, can be further
calculated as: 
\begin{equation}
C_{Q}\left[ \left\vert \psi _{S}\right\rangle ,\Pi _{l}\left( \phi \right) %
\right] =4[\left\langle \Psi \right\vert H_{1}\left\vert \Psi \right\rangle
-\left\vert \left\langle \Psi \right\vert H_{2}\left\vert \Psi \right\rangle
\right\vert ^{2}],  \label{13}
\end{equation}%
where $\left\vert \Psi \right\rangle $ is the state before loss and after
the first non-Gaussian operation, and$\ H_{1}$ and $H_{2}$\ are defined as: 
\begin{eqnarray}
H_{1} &=&\overset{\infty }{\underset{l}{\sum }}\frac{d\Pi _{l}^{\dag }\left(
\phi ,\eta ,\lambda \right) }{d\phi }\frac{d\Pi _{l}\left( \phi ,\eta
,\lambda \right) }{d\phi },  \label{14} \\
H_{2} &=&\overset{\infty }{\underset{l}{i\sum }}\frac{d\Pi _{l}^{\dag
}\left( \phi ,\eta ,\lambda \right) }{d\phi }\Pi _{l}\left( \phi ,\eta
,\lambda \right) .  \label{15}
\end{eqnarray}
The Kraus operation $\Pi _{l}\left( \phi ,\eta ,\lambda \right) $ can be
chosen as: 
\begin{equation}
\Pi _{l}\left( \phi ,\eta ,\lambda \right) =\sqrt{\frac{\left( 1-\eta
\right) ^{l}}{l!}}e^{i\phi \left( n-\lambda l\right) }\eta ^{\frac{n}{2}%
}a^{l},  \label{16}
\end{equation}%
where $n$ is the photon number operator, $\lambda =0$ and $\lambda =-1$
represent that the loss occur before and after the phase shift,
respectively. $\eta \ $quantifies the photon loss, where $\eta =1\ $and $%
\eta =0\ $represent the situations of complete lossy and absorption. The
parameters $\Pi _{l}\left( \phi ,\eta ,\lambda \right) $\ defines a set of
Kraus operators, which are used to minimize the value of $C_{Q}$. In our
model, the Fisher information under lossy case can be calculated as \textbf{%
\cite{a23}} 
\begin{equation}
F_{L}=\frac{4\eta \langle n\rangle \langle \Delta n^{2}\rangle }{\left(
1-\eta \right) \langle \Delta n^{2}\rangle +\eta \langle n\rangle },
\end{equation}%
where $\langle n\rangle $ and $\langle \Delta n^{2}\rangle \ $are the total
average photon number and the variance of the total average photon number of
mode $a$ inside the SU(1,1) interferometer. And it can be calculated as: 
\begin{equation}
\langle n\rangle =A^{2}\left\langle \Psi \right\vert n_{a}\left\vert \Psi
\right\rangle =A^{2}Q_{m,1,1,0,0},
\end{equation}%
and 
\begin{eqnarray}
\langle \Delta n^{2}\rangle &=&A^{2}\left\langle \Psi \right\vert
n_{a}^{2}\left\vert \Psi \right\rangle -\langle n\rangle  \notag \\
&=&4[A^{2}\left( Q_{m,2,2,0,0}+Q_{m,1,1,0,0}\right)  \notag \\
&&-\left( A^{2}Q_{m,1,1,0,0}\right) ^{2}].
\end{eqnarray}

In the presence of photon loss, we analyze the influence of each parameter
on the QFI. By observing the QFI varying with $\eta \ $and$\ \alpha \ $shown
in Fig. 9 and Fig. 10, it is clear to see that, (i) with the increasing of $%
\eta \ $and $\alpha $, QFI will become higher. (ii) Compared with the L-PSO,
the QFI of the D-PSO under lossy case can cover the advantages of L-PSO and
it even exceeds L-PSO when $m=1$, $\eta \ >0.6 $ and $\alpha >1.2$. This
trend will gradually become more obvious as order $m$, $\eta \ $ and \ $%
\alpha \ $increases. 
\begin{figure}[tbp]
\label{Figure9} \centering\includegraphics[width=0.95%
\columnwidth]{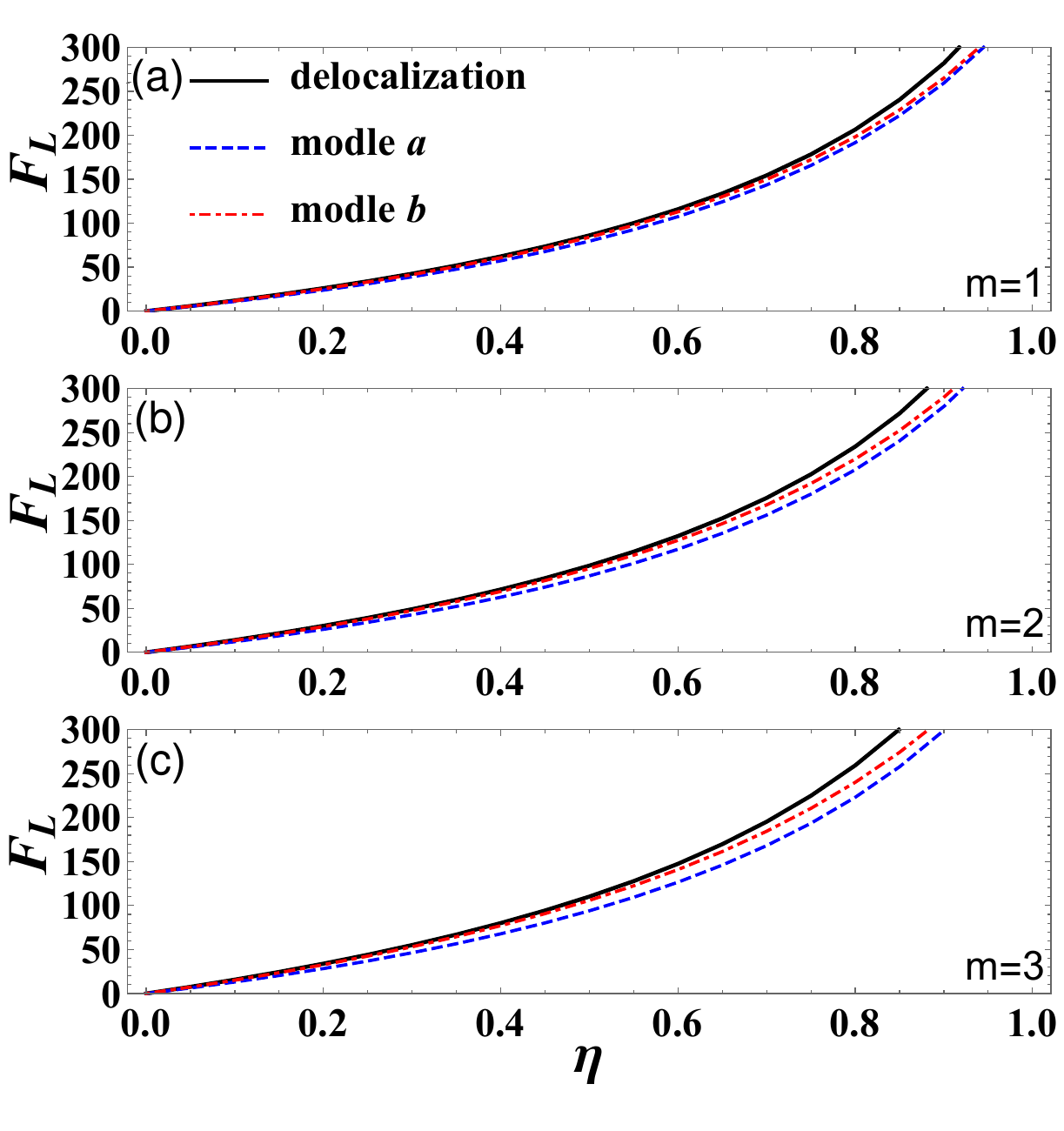} \ 
\caption{The comparison between D-PSO and L-PSO, on $F_{L}$ as a function of 
$\protect\eta $, with $g=1$, $\protect\alpha =3$.}
\end{figure}
\begin{figure}[tbp]
\label{Figure10} \centering\includegraphics[width=0.95%
\columnwidth]{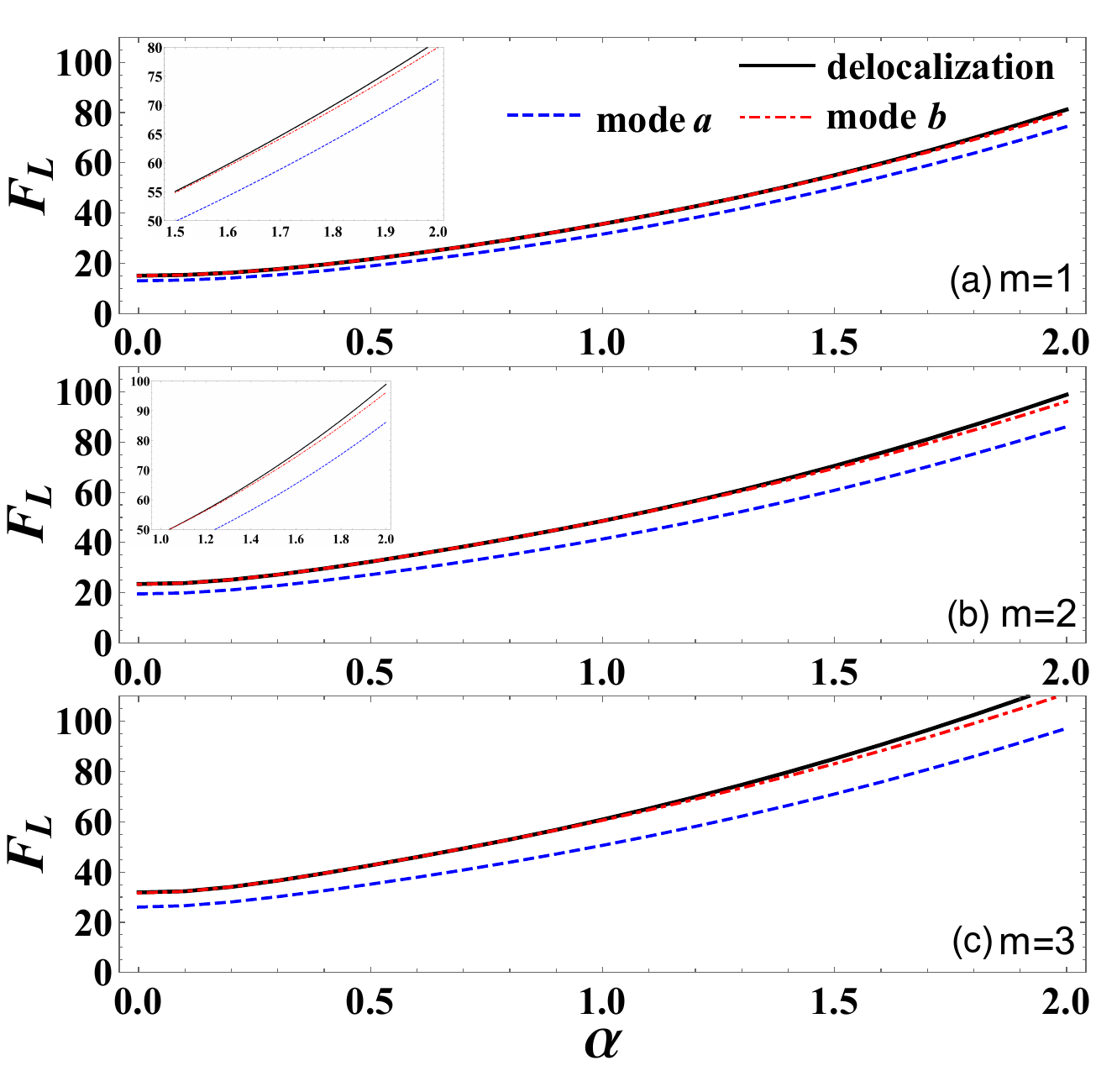} \ 
\caption{The comparison between D-PSO and L-PSO, on $F_{L}$ as a function of 
$\protect\alpha $, with $g=1$, $T=0.7$.}
\end{figure}
\begin{figure}[tph]
\label{Figure11} \centering\includegraphics[width=0.95%
\columnwidth]{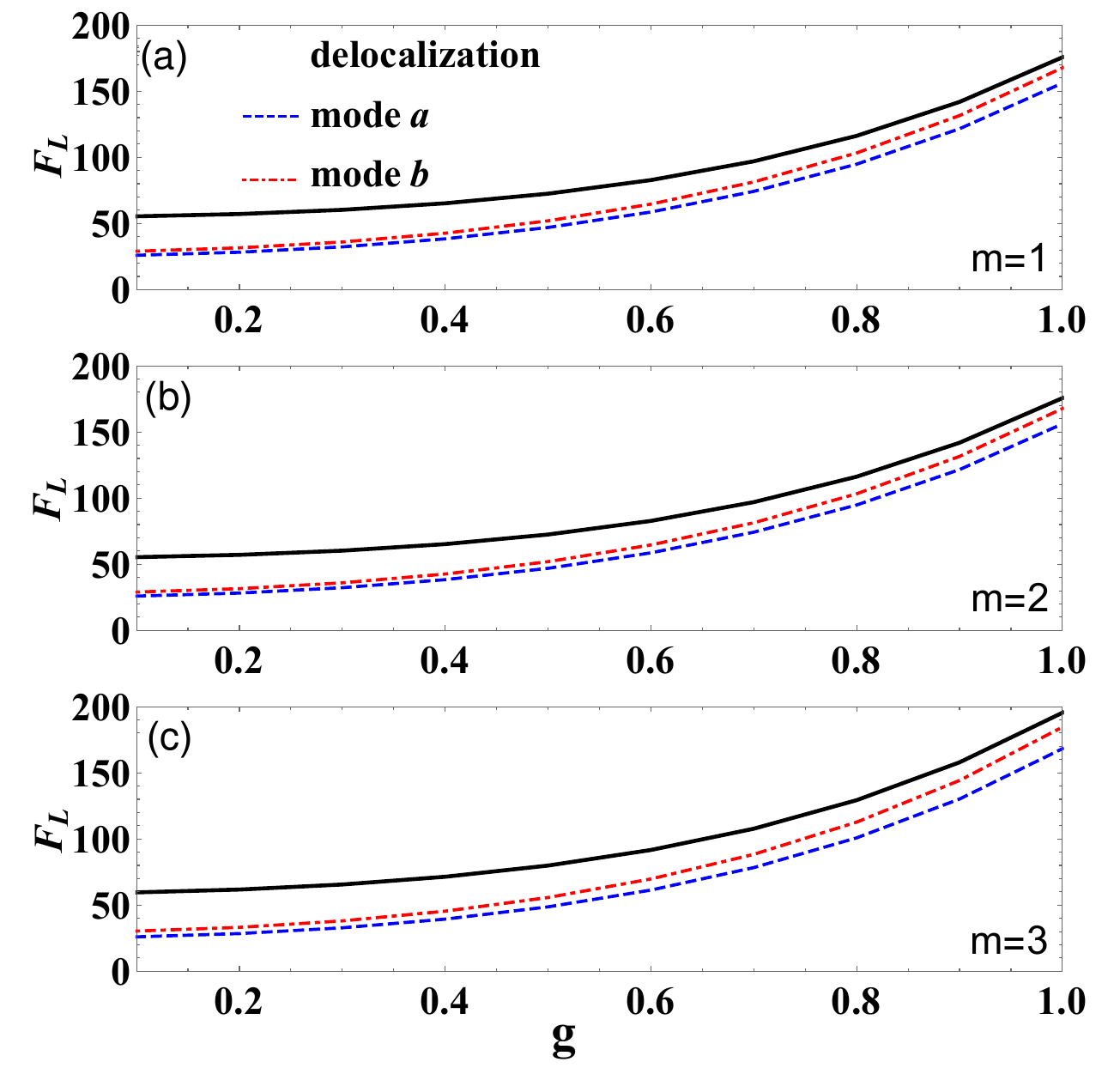} \ 
\caption{The comparison between D-PSO and L-PSO, on $F_{L}$ as a function of 
$g$, with $\protect\alpha =1$, $T=0.7$.}
\end{figure}

By observing the changes of the QFI with gain factor $g$ shown in Fig. 11,
it can be seen that, (i) the QFI in loss gradually increases with the
increase of the gain factor $g$. (ii) The QFI of L-PSO in mode $b$ in lossy
case exceeds that in mode $a$, and D-PSO can combine the advantages of both
L-PSO operations across the entire range of $g\ $and significantly
outperform their QFI. (iii) The trend that D-PSO is superior to L-PSO
increases with the increase of the order $m$, and the QFI differerence
becomes smaller as gain $g$.

\subsection{Comparison phase sensitivities and theoretical limits}

In this subsection, we compare the phase sensitivity with some theoretical
limits, including the QCRB, the SQL, and the HL. The QCRB is often used to
determine the ultimate phase precision of an interferometer and can be
derived by the QFI \textbf{\cite{a24}. }Therefore, in our scheme, the QCRB
in ideal case is given by: 
\begin{equation}
\Delta \phi _{QCRB}=\frac{1}{\sqrt{vF}},
\end{equation}%
where $v$ represents the number of measurements. For simplicity, we set $v=1$%
. The SQL and HL are defined as: $\Delta \phi _{SQL}=1/\sqrt{N}$ and $\Delta
\phi _{HL}=1/N$, where $N$ represent the total average photon number inside
the interferometer and before the second OPA. $N$ can be calculated as:%
\begin{eqnarray}
N &=&\left\langle \Psi _{\phi }\right\vert \left( a^{\dag }a+b^{\dag
}b\right) |\Psi _{\phi }\rangle  \notag \\
&=&A^{2}(Q_{1,1,0,0}+Q_{0,0,1,1}).
\end{eqnarray}

Figs. 12 and 13 show the comparison of the phase sensitivity $\Delta \phi $
for L-PSO and D-PSO with the SQL, the HL and the QCRB when $m=1$, $m=2$ and $%
m=3$ respectively, under the condition of $T=1$ and $T=0.7$. We can find in
Fig. 12, (i) as the order $m$ increases, the D-PSO can break through the
SQL, and it even can exceed the HL when $m>1\ $and $\alpha <1$. (ii) The
phase sensitivity of the D-PSO can get close to the QCRB when the order $m$
increases and is closer than that of the\ L-PSO. It's interesting to find
from Fig. 13 that, in a lossy case, (i) the D-PSO can still break through
the SQL, and exceed the HL when $m=1$ and $\alpha <0.3$. This trend will be
more obvious as the order $m$ grows. A comparison between Fig. 12 and Fig.
13 shows that both D-PSO and L-PSO are robust against internal loss.
However, when the order $m$ and coherent amplitude $\alpha $ increase, D-PSO
outperforms L-PSO. 
\begin{figure}[tbp]
\label{Figure12} \centering\includegraphics[width=0.95%
\columnwidth]{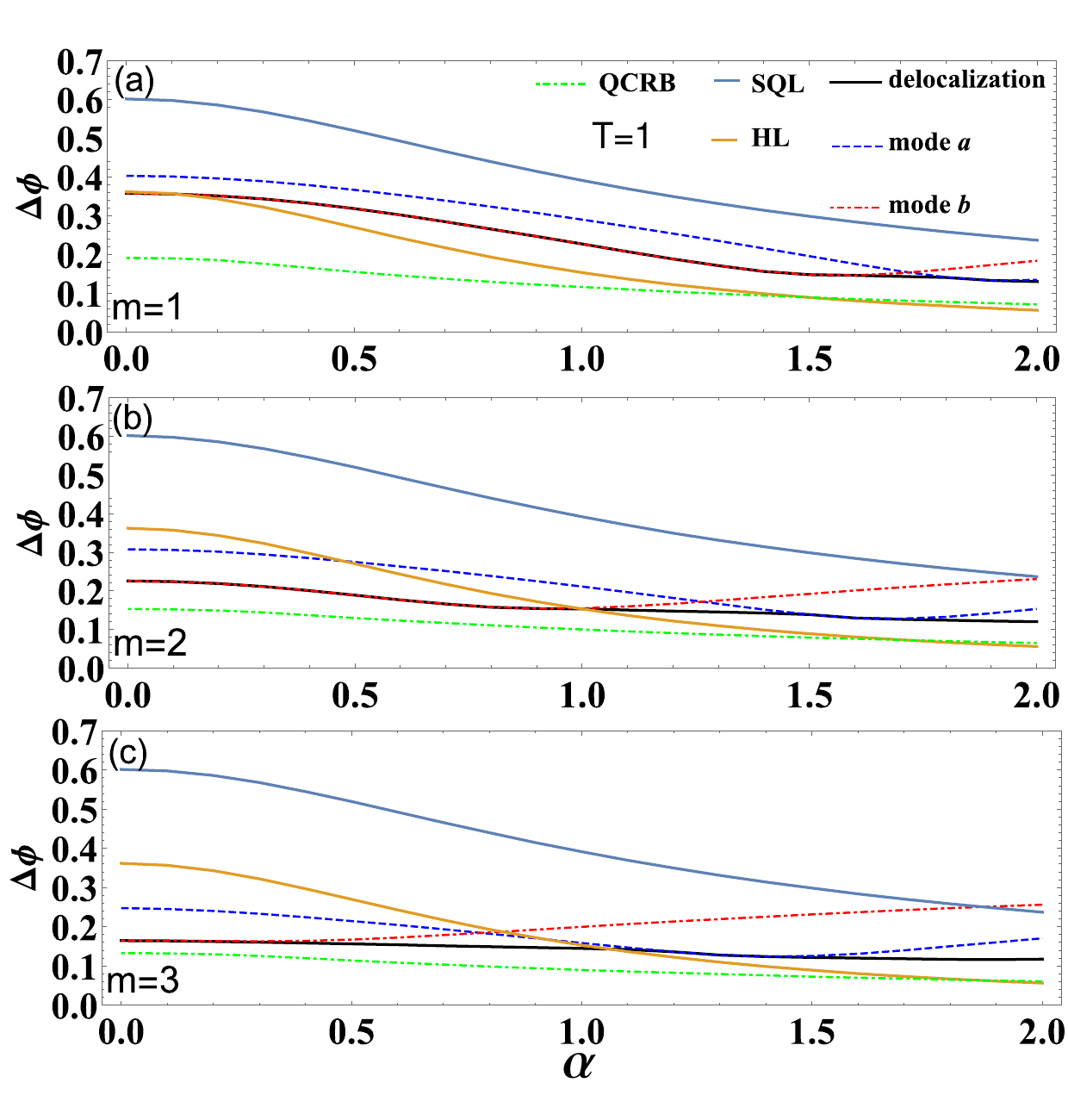} \ 
\caption{The comparison of the phase sensitivity $\Delta \protect\phi $ \ in
D-PSO and L-PSO with the SQL, the HL, and the QCRB, where gain factor $g=1$, 
$\protect\varphi =1\ $and $T=1$.}
\end{figure}
\begin{figure}[tbp]
\label{Figure13} \centering\includegraphics[width=0.95%
\columnwidth]{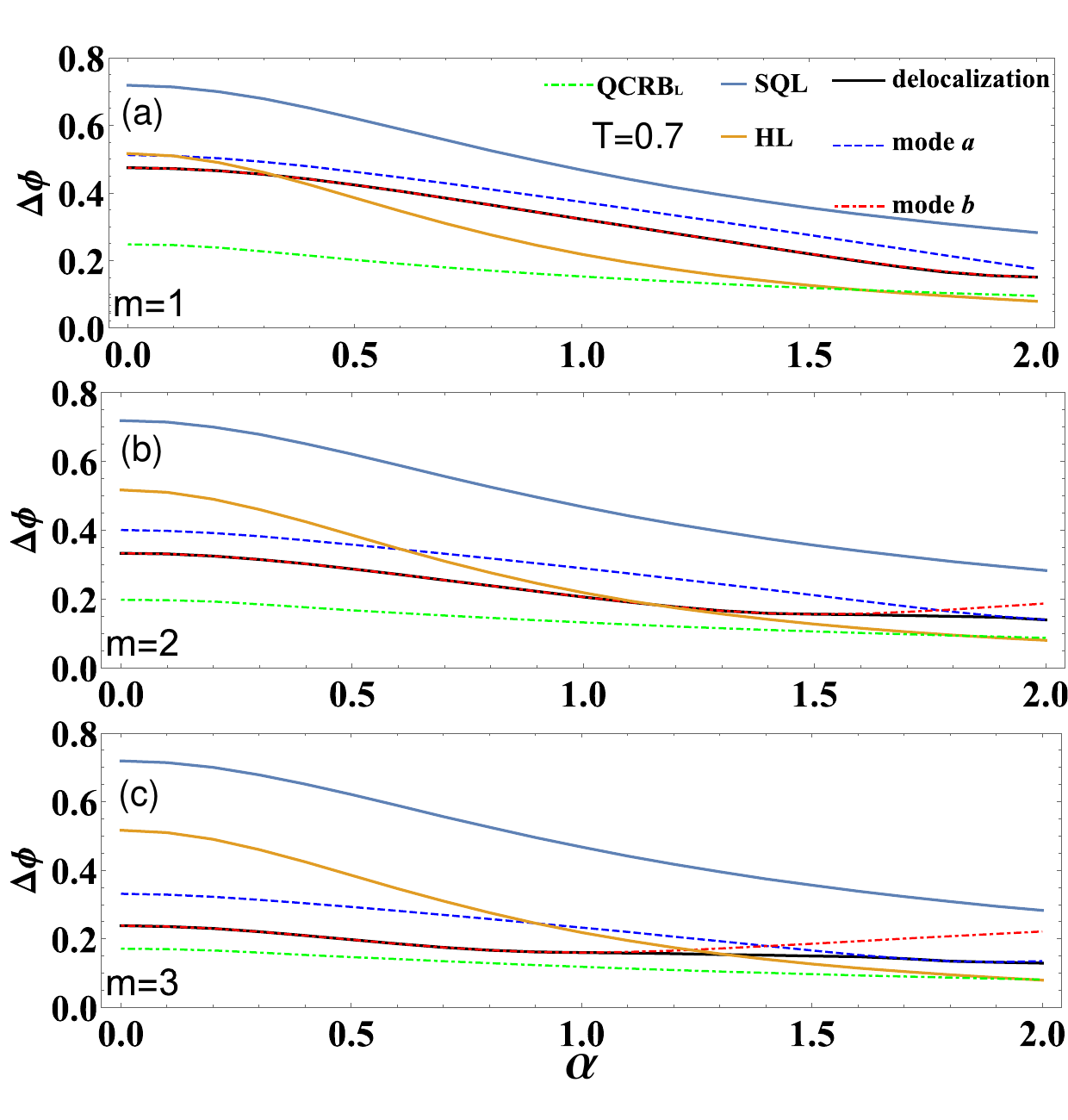} \ 
\caption{The comparison of the phase sensitivity $\Delta \protect\phi $ \ in
D-PSO and L-PSO with the SQL, the HL, and the QCRB, where gain factor $g=1$, 
$\protect\varphi =1\ $and $T=0.7$.}
\end{figure}

\section{Conclusion}

In this paper, we have studied the effects of implementing the D-PSO inside
the SU(1,1) interferometer on the phase sensitivity and QFI under both the
ideal and the photon-loss conditions. In addition, we have also studied the
effects of the gain coefficient $g$ of the OPA, the coherent amplitude $%
\alpha $, and the transmittance $T$ of the BS on the performance of the
system. Through analysis and comparison, we have verified that the D-PSO can
improve the measurement accuracy of the SU(1,1) interferometer and enhance
its robustness against internal photon loss. In addition, we also compared
the differences between the D-PSO and the L-PSO. In terms of phase
sensitivity, after optimization, the optimal range of phase shift $\varphi $%
, the coherent amplitude $\alpha $, the gain coefficient $g\ $and the
transmittance $T\ $in D-PSO can encompass the optimal ranges of both mode $a$
and mode $b$, which means that the D-PSO is able to combine and exceed the
advantages of the L-PSO. In terms of QFI, regardless of whether it is in an
ideal or realistic situation, compared with the L-PSO, the QFI of the D-PSO
will be larger under certain circumstances. Furthermore, as the order $m$
increases, the phase sensitivity of the D-PSO can get closer to the QCRB and
has a stronger ability to resist internal loss. Our results are hoped to
find important applications in quantum metrology and quantum information. 
\begin{acknowledgments}
This work is supported by the National Natural Science Foundation of China (Grants No. 11964013 and No. 12104195), the Jiangxi Provincial Natural Science Foundation (Grants No. 20242BAB26009 and 20232BAB211033), as well as the Jiangxi Provincial Key Laboratory of Advanced Electronic Materials and Devices (Grant No. 2024SSY03011), Jiangxi Civil-Military Integration Research Institute (Grant No. 2024JXRH0Y07), and the Science and Technology Project of Jiangxi Provincial Department of Science and Technology (Grant No. GJJ2404102).
\end{acknowledgments}\bigskip

\textbf{APPENDIX\ A : THE CALCULATION OF THE PHASE SENSITIVITY FOR THE D-PSO}

In this Appendix, we give the derivation of the phase sensitivity $\Delta
\phi $. In our scheme, the transform relation between the output state $%
\left\vert \Psi _{out}\right\rangle$ and the input state $\left\vert \Psi _{in}\right\rangle $is given by: 
\begin{equation}
\left\vert \Psi _{out}\right\rangle =AU_{S_{2}}U_{\phi }U_{B}\left(
sa+tb\right) ^{m}U_{S_{1}}\left\vert \Psi _{in}\right\rangle .  \tag{A1}
\end{equation}%
Before deriving the phase sensitivity (Eq. (3)), we introduce a formula,
i.e.,%
\begin{align}
& Q_{m,x_{1},y_{1},x_{2},y_{2}}  \notag \\
& =\left\langle \Psi _{in}\right\vert U_{S_{1}}^{\dagger }\left(
sa^{\dagger }+tb^{\dagger }\right) ^{m}U_{B}^{\dagger }\times \left(
a^{\dagger x_{1}}a^{y_{1}}b^{\dagger x_{2}}b^{y_{2}}\right)   \notag \\
& \times U_{B}\left( sa+tb\right) ^{m}U_{S_{1}}\left\vert \Psi
_{in}\right\rangle   \notag \\
& =\frac{\partial ^{2m+x_{1}+y_{1}+x_{2}+y_{2}}}{\partial \lambda
_{3}^{m}\partial \lambda _{4}^{m}\partial \lambda _{5}^{x_{1}}\partial
\lambda _{6}^{y_{1}}\partial \lambda _{7}^{x_{2}}\partial \lambda
_{8}^{y_{2}}}  \notag \\
& \times U_{S_{1}}^{\dagger }e^{\lambda _{3}s^{\ast }a^{\dag }}e^{\lambda
_{3}t^{\ast }b^{\dag }}U_{B}^{\dagger }e^{\lambda _{5}a^{\dag }}e^{\lambda
_{6}a}e^{\lambda _{7}b^{\dag }}e^{\lambda _{8}b}U_{B}  \notag \\
& \times e^{s\lambda _{4}a}e^{\lambda _{4}tb}U_{S1}|_{\lambda _{3}=\lambda
_{4}=\lambda _{5}=\lambda _{6}=\lambda _{7}=\lambda _{8}=0}  \notag \\
& =\frac{\partial ^{2m+x_{1}+y_{1}+x_{2}+y_{2}}}{\partial \lambda
_{3}^{m}\partial \lambda _{4}^{m}\partial \lambda _{5}^{x_{1}}\partial
\lambda _{6}^{y_{1}}\partial \lambda _{7}^{x_{2}}\partial \lambda
_{8}^{y_{2}}}  \notag \\
& \times \left\{ \exp \left[ w_{4}\right] \right\} _{_{\lambda _{3}=\lambda
_{4}=\lambda _{5}=\lambda _{6}=\lambda _{7}=\lambda _{8}=0,}}  \tag{A2}
\end{align}%
where 
\begin{align}
w_{1}& =-(\lambda _{8}\sqrt{T}+\lambda _{4}t)\sinh g  \notag \\
& \times \lbrack (\lambda _{6}\sqrt{T}+\lambda _{4}s)\cosh g  \notag \\
& -(\lambda _{7}\sqrt{T}+\lambda _{3})\sinh g]  \notag \\
& -(\lambda _{5}\sqrt{T}+\lambda _{3})\sinh g  \notag \\
& \times \lbrack (\lambda _{7}\sqrt{T}+\lambda _{3})\cosh g  \notag \\
& -(\lambda _{6}\sqrt{T}+\lambda _{4}s)\sinh g],  \tag{A3}
\end{align}%
\begin{align}
w_{2}& =(\lambda _{5}\sqrt{T}+\lambda _{3})\cosh g  \notag \\
& -(\lambda _{8}\sqrt{T}+\lambda _{4}t)\sinh g,  \tag{A4}
\end{align}%
\begin{align}
w_{3}& =(\lambda _{6}\sqrt{T}+\lambda _{4}s)\cosh g  \notag \\
& -(\lambda _{7}\sqrt{T}+\lambda _{3})\sinh g,  \tag{A5}
\end{align}%
\begin{equation}
w_{4}=w_{1}+w_{2}\alpha ^{\ast }+w_{3}\alpha .  \tag{A6}
\end{equation}%
Here $m$, $x_{1}$, $x_{2}$, $y_{1}$, $y_{2}$ are positive integers, $\lambda
_{3}$, $\lambda _{4}$, $\lambda _{5}$, $\lambda _{6}$, $\lambda _{7}$, $%
\lambda _{8}$ are differential variables. After differentiation, all these
differential variables take zero.

In order to derive Eq. (3), we use the transformation relations, i.e. 
\begin{align}
U_{S1}^{\dagger }aU_{S1}& =a\cosh g-b^{\dagger }e^{i\theta _{1}}\sinh g,\  
\notag \\
U_{S1}^{\dagger }bU_{S1}& =b\cosh g-a^{\dagger }e^{i\theta _{1}}\sinh g,\  
\tag{A7}
\end{align}%
\begin{align}
U_{S_{2}}^{\dagger }aU_{S2}& =a\cosh g-b^{\dagger }e^{i\theta _{2}}\sinh g, 
\notag \\
U_{S2}^{\dagger }bU_{S2}& =b\cosh g-a^{\dagger }e^{i\theta _{2}}\sinh g,\  
\tag{A8}
\end{align}%
\begin{align}
U_{B}^{\dagger }aU_{B}& =\sqrt{T}a+\sqrt{R}a_{\nu },  \notag \\
U_{B}^{\dagger }a^{\dagger }U_{B}& =\sqrt{T}a^{\dagger }+\sqrt{R}a_{\nu
}^{\dagger },  \tag{A9}
\end{align}%
\begin{align}
U_{B}^{\dagger }bU_{B}& =\sqrt{T}b+\sqrt{R}b_{v},  \notag \\
U_{B}^{\dagger }b^{\dag }U_{B}& =\sqrt{T}b^{\dag }+\sqrt{R}b_{v}^{\dag }. 
\tag{A10}
\end{align}

In our scheme, the phase sensitivity can be calculated as: 

\begin{equation}
\Delta \phi =\frac{\sqrt{%
\begin{array}{c}
\left\langle \Psi _{out}\right\vert \left( a^{\dagger }a+b^{\dag }b\right)
^{2}\left\vert \Psi _{out}\right\rangle \\ 
-\left\langle \Psi _{out}\right\vert \left( a^{\dagger }a+b^{\dag }b\right)
\left\vert \Psi _{out}\right\rangle ^{2}%
\end{array}%
}}{|\partial \left\langle \Psi _{out}\right\vert \left( a^{\dagger
}a+b^{\dag }b\right) \left\vert \Psi _{out}\right\rangle /\partial \phi |}, 
\tag{A11}
\end{equation}%
where 
\begin{align}
& \left\langle \Psi _{out}\right\vert \left( a^{\dagger }a+b^{\dag }b\right)
^{2}\left\vert \Psi _{out}\right\rangle  \notag \\
& =\left\langle \Psi _{out}\right\vert \left( a^{\dagger 2}a^{2}\right)
\left\vert \Psi _{out}\right\rangle +\left\langle \Psi _{out}\right\vert
\left( a^{\dagger }a\right) \left\vert \Psi _{out}\right\rangle  \notag \\
& +2\left\langle \Psi _{out}\right\vert \left( a^{\dagger }ab^{\dag
}b\right) \left\vert \Psi _{out}\right\rangle  \notag \\
& +\left\langle \Psi _{out}\right\vert \left( b^{\dagger 2}b^{2}\right)
\left\vert \Psi _{out}\right\rangle +\left\langle \Psi _{out}\right\vert
\left( b^{\dagger }b\right) \left\vert \Psi _{out}\right\rangle ,  \tag{A12}
\end{align}%
and
\begin{align}
& \left\langle \Psi _{out}\right\vert \left( a^{\dagger 2}a^{2}\right)
\left\vert \Psi _{out}\right\rangle  \notag \\
& =\cosh ^{4}gQ_{m,2,2,0,0}+2e^{-i\phi }\sinh g\cosh ^{3}gQ_{m,2,1,1,0} 
\notag \\
& +e^{-2i\phi }\sinh ^{2}g\cosh ^{2}gQ_{m,2,0,2,0}  \notag \\
& +2e^{i\phi }\sinh g\cosh ^{3}gQ_{m,1,2,0,1}  \notag \\
& +4\sinh ^{2}g\cosh ^{2}g\left( Q_{m,1,1,1,1}+Q_{m,1,1,0,0}\right)  \notag
\\
& +2e^{-i\phi }\sinh ^{3}g\cosh g\left( Q_{m,1,0,2,1}+2Q_{m,1,0,1,0}\right) 
\notag \\
& +e^{2i\phi }\sinh ^{2}g\cosh ^{2}gQ_{m,0,2,0,2}  \notag \\
& +2e^{i\phi }\sinh ^{3}g\cosh g\left( Q_{m,0,1,1,2}+2Q_{m,0,1,0,1}\right) 
\notag \\
& +\sinh ^{4}g\left( Q_{m,0,0,2,2}+4Q_{m,0,0,1,1}\right)  \notag \\
& +2\sinh ^{4}gQ_{m,0,0,0,0},  \tag{A13}
\end{align}%
\begin{align}
& \left\langle \Psi _{out}\right\vert \left( a^{\dagger }a\right) \left\vert
\Psi _{out}\right\rangle  \notag \\
& =\cosh ^{2}gQ_{m,1,1,0,0}+e^{-i\phi }\sinh g\cosh gQ_{m,1,0,1,0}  \notag \\
& +e^{i\phi }\sinh g\cosh gQ_{m,0,1,0,1}  \notag \\
& +\left( Q_{m,0,0,1,1}+Q_{m,0,0,0,0}\right) \sinh ^{2}g,  \tag{A14}
\end{align}%
and
\begin{align}
& \left\langle \Psi _{out}\right\vert \left( b^{\dagger 2}b^{2}\right)
\left\vert \Psi _{out}\right\rangle  \notag \\
& =\cosh ^{4}gQ_{m,0,0,2,2}+2e^{-i\phi }\sinh g\cosh ^{3}gQ_{m,1,0,2,1} 
\notag \\
& +e^{-2i\phi }\sinh ^{2}g\cosh ^{2}gQ_{m,2,0,2,0}  \notag \\
& +2e^{i\phi }\sinh g\cosh ^{3}gQ_{m,0,1,1,2}  \notag \\
& +4\sinh ^{2}g\cosh ^{2}g\left( Q_{m,1,1,1,1}+Q_{m,0,0,1,1}\right)  \notag
\\
& +2e^{-i\phi }\sinh ^{3}g\cosh g\left( Q_{m,2,1,1,0}+2Q_{m,1,0,1,0}\right) 
\notag \\
& +e^{2i\phi }\sinh ^{2}g\cosh ^{2}gQ_{m,0,2,0,2}  \notag \\
& +2e^{i\phi }\sinh ^{3}g\cosh g\left( Q_{m,1,2,0,1}+2Q_{m,0,1,0,1}\right) 
\notag \\
& +\sinh ^{4}g\left( Q_{m,2,2,0,0}+4Q_{m,1,1,0,0}\right)  \notag \\
& +2\sinh ^{4}gQ_{m,0,0,0,0},  \tag{A15}
\end{align}%

\begin{align}
& \left\langle \Psi _{out}\right\vert \left( b^{\dagger }b\right) \left\vert
\Psi _{out}\right\rangle  \notag \\
& =\cosh ^{2}gQ_{m,0,0,1,1}+e^{-i\phi }\sinh g\cosh gQ_{m,1,0,1,0}  \notag \\
& +e^{i\phi }\sinh g\cosh gQ_{m,0,1,0,1}  \notag \\
& +\sinh ^{2}gQ_{m,1,1,0,0}+\sinh ^{2}gQ_{m,0,0,0,0}.  \tag{A16}
\end{align}%


\bigskip

\begin{thebibliography}{99}
\bibitem{c1} Giovannetti, V., Lloyd, S. \& Maccone, L. Advances in quantum
metrology. Nature Photon 5, 222--229 (2011).

\bibitem{c2} Gregory Krueper, Lior Cohen, and Juliet T. Gopinath, Cascaded
multiparameter quantum metrology, Phys. Rev. A. 111, 012618 (2025).

\bibitem{c3} S. Pang and A. N. Jordan, Optimal adaptive control for quantum
metrology with time-dependent Hamiltonians, Nat. Commun. 8, 14695 (2017).

\bibitem{c4} W. Ge, K. Jacobs, Z. Eldredge, A. V. Gorshkov, and M.
Foss-Feig, Distributed Quantum Metrology with Linear Networks and Separable
Inputs, Phys. Rev. Lett. 121, 043604 (2018).

\bibitem{c5} S. M. Roy and S. L. Braunstein, Exponentially enhanced quantum
metrology, Phys. Rev. Lett. 100, 220501 (2008).

\bibitem{c6} T. Corbitt and N. Mavalvala, Quantum noise in
gravitational-wave interferometers, Journal of optics. B, S675--S683 (2004).

\bibitem{c7} R. X. Adhikari, Gravitational radiation detection with laser
interferometry, Rev. Mod. Phys. 86, 121 (2014).

\bibitem{c8} E. Oelker, L. Barsotti, S. Dwyer, D. Sigg, and N. Mavalvala,
Squeezed light for advanced gravitational wave detectors and beyond, Opt.
Express, 22, 21106--21121 (2014).

\bibitem{c9} H.Vahlbruch, D. Wilken, M. Mehmet, and B. Willke, Laser power
stabilization beyond the shot noise limit using squeezed light, Phys. Rev.
L. 121, 173601 (2018).

\bibitem{c10} P. Gill, Atomic clocks- raising the standards, Sci. 294,
1666--1668 (2001).

\bibitem{c11} S. Weinberg, Lindblad decoherence in atomic clocks, Phys. Rev.
A. 94, 042117 (2016).

\bibitem{c12} M. Arndt and C. Brand, Interference of atomic clocks, Sci.
349, 1168--1169 (2015).

\bibitem{c13} Andrew D. Ludlow, Martin M. Boyd, Jun Ye, E. Peik, and
P.\thinspace O. Schmidt, Optical atomic clocks, Rev. Mod. Phys. 87, 637
(2015).

\bibitem{c14} N. Bornman, S. Prabhakar, A. Valles, J. Leach, and A. Forbes,
Ghost imaging with engineered quantum states by hong-ou-mandel interference,
New J. Phys 21, 073044 (2019).

\bibitem{c15} M. Tsang, Quantum imaging beyond the diffraction limit by
optical centroid measurements, Phys. Rev. L. 102, 253601 (2009).

\bibitem{c16} C. Thiel, T. Bastin, J. Martin, E. Solano, J. von Zanthier,
and G. S. Agarwal, Quantum imaging with incoherent photons, Phys. Rev. L.
99, 133603 (2007).

\bibitem{c17} A. Mikhalychev, S. Almazrouei, S. Mikhalycheva, A. Bouchalkha,
D. Mogilevtsev, and B. Ahmedov, Efficient estimation of error bounds for
quantum multiparametric imaging with constraints, Phys. Rev. A. 111, 032618
(2025).

\bibitem{c18} S. D. Huver, C. F. Wildfeuer, and J. P. Dowling, Entangled
fock states for robust quantum optical metrology, imaging, and sensing,
Phys. Rev. A. 78, 063828 (2008).

\bibitem{a1} Jianxin Chen, Toby S. Cubitt, Aram W. Harrow, and Graeme Smith,
Entanglement can Completely Defeat Quantum Noise, Phys. Rev. Lett. 107,
250504 (2011).

\bibitem{a2} C. M. Caves, Quantum-mechanical noise in an interferometer,
Phys. Rev. D 23, 1693 (1981).

\bibitem{a3} Dowling, J. P. Quantum optical metrology -- the lowdown on
high-N00N states. Contemporary Physics, 49(2), 125--143 (2008).

\bibitem{a4} T. Kim, J. Shin, Y. Ha, H. Kim, G. Park, T. G. Noh, and C. K.
Hong, The phase-sensitivity of a Mach--Zehnder interferometer for Fock state
inputs, Opt. Commun. 156, 37 (1998).

\bibitem{a5} Luca Pezz\'{e} and Augusto Smerzi, Ultrasensitive Two-Mode
Interferometry with Single-Mode Number Squeezing, Phys. Rev. Lett. 110,
163604 (2013).

\bibitem{b1} Luca Pezz\'{e} and Augusto Smerzi, Mach-Zehnder Interferometry
at the Heisenberg Limit with Coherent and Squeezed-Vacuum Light, Phys. Rev.
Lett. 100, 073601 (2008).

\bibitem{a6} Stefan Ataman, Optimal Mach-Zehnder phase sensitivity with
Gaussian states, Phys. Rev. A. 100, 063821 (2019).

\bibitem{a7} Karunesh K. Mishra, and Stefan Ataman, Optimal phase
sensitivity of an unbalanced Mach-Zehnder interferometer, Phys. Rev. A. 106,
023716 (2022).

\bibitem{b2} Zekun Zhao, Qingqian Kang, Huan Zhang, Teng Zhao, Cunjin Liu,
and Liyun Hu, Phase estimation via coherent and photon-catalyzed squeezed
vacuum states, Opt. Express 32, 28267 (2024).

\bibitem{a8} B. Yurke, S. L. McCall, and J. R. Klauder, SU (2) and SU (1,1)
interferometers, Phys. Rev. A 33(6), 4033--4054 (1986).

\bibitem{a9} Jian-Dong Zhang, Chenglong You, Chuang Li, and Shuai Wang,
Phase sensitivity approaching the quantum Cram\'{e}r-Rao bound in a modified
SU(1,1) interferometer, Phys. Rev. A. 103,032617 (2021).

\bibitem{a10} Huan Zhang, Wei Ye, Chaoping Wei, Ying Xia, Shoukang Chang,
Zeyang Liao, and Liyun Hu, Improved phase sensitivity in a quantum optical
interferometer based on multiphoton catalytic two-mode squeezed vacuum
states, Phys. Rev. A. 103, 013705 (2021).

\bibitem{a11} Qingqian Kang, Zekun Zhao, Teng Zhao, Cunjin Liu, and Liyun
Hu, Phase estimation via a number-conserving operation inside a SU(1,1)
interferometer, Phys. Rev. A. 110, 022432 (2024).

\bibitem{a12} Kun Zhang, Yinghui Lv, Yu Guo, Jietai Jing, and Wu-Ming Liu,
Enhancing the precision of a phase measurement through phase-sensitive
non-Gaussianity, Phys. Rev, A. 105, 042607 (2022).

\bibitem{b3} Mattia Walschaers, Non-Gaussian Quantum States and Where to
Find Them, PRX. Quantum. 2, 030204 (2021).

\bibitem{b4} L. L. Guo, Y. F. Yu, and Z. M. Zhang, Improving the phase
sensitivity of an SU(1,1) interferometer with photon-added squeezed vacuum
light, Opt. Express 26, 29099 (2018).

\bibitem{b5} Y. Ouyang, S. Wang, and L. J. Zhang, Quantum optical
interferometry via the photon-added two-mode squeezed vacuum states, J. Opt.
Soc. Am. B 33, 1373 (2016).

\bibitem{b6} R. Birrittella and C. C. Gerry, Quantum optical interferometry
via the mixing of coherent and photon-subtracted squeezed vacuum states of
light, J. Opt. Soc. Am. B 31, 586 (2014).

\bibitem{b7} Youke Xu, Teng Zhao, Qingqian Kang, Cunjin Liu, Liyun Hu, and
Sanqiu Liu, Phase sensitivity of an SU(1,1) interferometer in photon-loss
via photon operations, Opt. Express, 31, 8414 (2023).

\bibitem{b8} S. K. Chang, W. Ye, H. Zhang, L. Y. Hu, J. H. Huang, and S. Q.
Liu, Improvement of phase sensitivity in an SU(1,1) interferometer via a
phase shift induced by a Kerr medium, Phys. Rev. A. 105, 033704 (2022).

\bibitem{b9} Qingqing Sun, Hyunchul Nha, and M. Suhail Zubairy, Entanglement
criteria and nonlocality for multimode continuous-variable systems, Phys.
Rev. A. 80.020101 (2009).

\bibitem{a13} Nicola Biagi, Luca S. Costanzo, Marco Bellini, and Alessandro
Zavatta, Generating Discorrelated States for Quantum Information Protocols
by Coherent Multimode Photon Addition, Adv. Quantum. Tech. 4, 2000141 (2021).

\bibitem{a15} Nicola Biagi, Saverio Francesconi, Manuel Gessner, Marco
Bellini, and Alessandro Zavatta, Remote Phase Sensing by Coherent Single
Photon Addition, Adv. Quantum. Technol. 5, 2200039 (2022).

\bibitem{f1} Nicola Biagi, Luca S. Costanzo, Marco Bellini, and Alessandro
Zavatta, Entangling Macroscopic Light States by Delocalized Photon Addition,
Phys. Rev. Lett. 124, 033604 (2020).

\bibitem{a16} Stefan Ataman, Phase sensitivity of a Mach-Zehnder
interferometer with single-intensity and difference-intensity detection,
Phys. Rev. A. 98, 043856 (2018).

\bibitem{a17} Shengshuai Liu, Yanbo Lou, Jun Xin, and Jietai Jing, Quantum
Enhancement of Phase Sensitivity for the Bright-Seeded SU(1,1)
Interferometer with Direct Intensity Detection, Phys. Rev. Applied. 10,
064046 (2018).

\bibitem{a18} D. Li, C. H. Yuan, Z. Y. Ou, and W. Zhang, The phase
sensitivity of an SU(1,1) interferometer with coherent and squeezed-vacuum
light, New J. Phys. 16, 073020 (2014).

\bibitem{a19} X. Y. Hu, C. P. Wei, Y. F. Yu, and Z. M. Zhang, Enhanced phase
sensitivity of an SU(1,1) interferometer with displaced squeezed vacuum
light, Front. Phys. 11, 114203 (2016).

\bibitem{a20} P. M. Anisimov, G. M. Raterman, A. Chiruvelli, W. N. Plick, S.
D. Huver, H. Lee, and J. P. Dowling, Quantum metrology with two-mode
squeezed vacuum: Parity detection beats the Heisenberg limit, Phys.Rev.Lett.
104, 103602 (2010).

\bibitem{a21} D. Li, B. T. Gard, Y. Gao, C. H. Yuan, W. Zhang, H. Lee, and
J. P. Dowling, Phase sensitivity at the Heisenberg limit in an SU(1,1)
interferometer via parity detection, Phys. Rev. A. 94, 063840 (2016).

\bibitem{a22} Marcin Jarzyna and Rafa\l\ Demkowicz-Dobrza\'{n}ski, Quantum
interferometry with and without an external phase reference, Phys. Rev. A.
85, 01180 (2012).

\bibitem{a23} B. M. Escher, R. L. de Matos Filho, and L. Davidovich, General
framework for estimating the ultimate precision limit in noisy
quantum-enhanced metrology, Nat. Phys. 7, 406 (2011).

\bibitem{a24} S. L. Braunstein and C. M. Caves, Statistical distance and the
geometry of quantum states, Phys. Rev. Lett. 72(22), 3439--3443 (1994).
\end{thebibliography}
\end{document}